\documentclass[prl,showkeys,twocolumn]{revtex4}
\usepackage{graphicx,amssymb,amsmath}
\usepackage{hyperref}
\usepackage{enumitem}
\usepackage{subfig}
\usepackage{empheq}
\usepackage{color,soul}
\usepackage{mathrsfs}
\usepackage{afterpage}

\begin{document}

\title{From global scaling to the dynamics of individual cities}

\author{Jules Depersin}
\affiliation{Institut de Physique Th\'eorique, Universit\'e Paris Saclay, CEA, CNRS, F-91191 Gif-sur-Yvette, France}

\author{Marc Barthelemy}
\email{marc.barthelemy@ipht.fr}
\affiliation{Institut de Physique Th\'eorique, Universit\'e Paris Saclay, CEA, CNRS, F-91191 Gif-sur-Yvette, France}
\affiliation{Centre d'Analyse et de Math\'ematique Sociales, (CNRS/EHESS) 54, Boulevard Raspail, 75006 Paris, France}

\begin{abstract}

Scaling has been proposed as a powerful tool to analyze the properties of complex systems, and in particular for cities where it describes how various properties change with population. The empirical study of scaling on a wide range of urban datasets displays apparent nonlinear behaviors whose statistical validity and meaning were recently the focus of many debates. We discuss here another aspect which is the implication of such scaling forms on individual cities and how they can be used for predicting the behavior of a city when its population changes. We illustrate this discussion on the case of delay due to traffic congestion with a dataset for 101 US cities in the range 1982-2014. We show that the scaling form obtained by agglomerating all the available data for different cities and for different years displays indeed a nonlinear behavior, but which appears to be unrelated to the dynamics of individual cities when their population grow. In other words, the congestion induced delay in a given city does not depend on its population only, but also on its previous history. This strong path-dependency prohibits the existence of a simple scaling form valid for all cities and shows that we cannot always agglomerate the data for many different systems. More generally, these results also challenge the use of transversal data for understanding longitudinal series for cities.

\end{abstract}


\maketitle

The recent availability of data for cities opens the fascinating possibility of a science of cities \cite{Batty:2013,Barthelemy:2016} and has led numerous scientists to search for general laws \cite{Pumain:2004,Bettencourt:2007} ruling the evolution of various socio-economical and structural indicators such as patent production, personal income or electric cable total length, etc. In \cite{Pumain:2004}, it was suggested that assuming the population $P$ to be the most important determinant for cities, we could study the evolution of many different features when $P$ is increasing. In \cite{Bettencourt:2007}, many socio-economic factors were studied versus population indicating the existence of simple scaling laws under the form of power laws. For each indicator $Y$, Bettencourt et al.~\cite{Bettencourt:2007} found a power law of the form $Y\sim P^\beta$ where the exponent $\beta$ depends on the quantity considered. Some quantities evolve \emph{superlinearly} with the population ($\beta >1$), for instance new patents ($\beta =1.27$), gross domestic product (GDP) ($1.13<\beta <1.26$) or serious crime ($\beta =1.16$), while some other behave \emph{sublinearly} ($\beta <1$) as gasoline stations or sales. Quantities that are independent from the size of the city -- typically human-related quantities such as water consumption -- scale with an exponent $\beta=1$. The usual explanation for these effects is the impact of interactions (scaling as $P^2$) for superlinear quantities, and economies of scale for sublinear quantities. This publication \cite{Bettencourt:2007} was followed by a wealth of other measures such as the abundance of business categories \cite{Youn:2016}, the number of sexually transmitted infection \cite{Petterson:2015}, road networks \cite{Samaniego:2008}, or carbon dioxide emissions \cite{Glaeser:2010,Fragkias:2013,Louf:2014a,Olive:2014,Rybski:2017}.

Scaling in urban systems has however been criticized in some recent papers \cite{Shalizi:2011,Arcaute:2015,Leitao:2016,Cottineau:2017,Louf:2014a}. A first re-analysis of the data for the GDP and income \cite{Shalizi:2011} showed that the power law could not be distinguished from other functional forms, or that the linear fit is better \cite{Arcaute:2015}, and in \cite{Leitao:2016} the authors led a rigorous investigation on the statistical quality of scalings for various quantities and found that in many superlinear cases, the linear assumption could in fact not be rejected. They also showed that the fitting results depend crucially on the assumptions about noise. From another point of view, the authors in \cite{Cottineau:2017} showed that, for some socioeconomic indicators, those scaling are not universal and could depend on details of urban systems. More precisely, they showed on data of $5,000$ french cities that two different definitions of the cities (\textit{Unit\'{e} urbaine} (Urban Units) and \textit{Aire urbaine} (Metropolitan areas)) lead to different values of the scaling exponent for a given quantity, a result confirmed on transport-emitted $CO_2$ in \cite{Louf:2014a}. Not only the value of the exponent can change, but in some case, for different definitions of the city, the scaling regime changes: for instance, the number of jobs in the manufacturing sector grows superlinearly with the population of Urban Units, but sublinearly if one considers Metropolitan Areas \cite{Cottineau:2017}. We can expect the results to change quantitatively, but here we have changes from the superlinear to the sublinear regime, casting some doubts about this nonlinear scaling and its universality.

In this paper we raise another problem that is the relevance of such a scaling for the individual dynamics of cities. At a more theoretical level, we question here the scaling assumption where a quantity $Y$ (usually extensive) is assumed to be determined by the population only $Y=F(P)$ (where $F$ is in general an unknown function). Even if the population is an important determinant for cities we cannot exclude time effect and path-dependency which would then imply that the quantity $Y$ depends also on time $Y=F(P,t)$ and possibly on all $Y(t')$ for $t'<t$. In other terms, the path-dependency means that it doesn't make sense in general to compare two cities having the same population but at very different dates: both central Paris and Phoenix (AZ) had a population of about 1 million inhabitants, the former in 1840 and the latter in 1990, and it is very likely that the dynamics -- for most of the relevant quantities -- from 1840 in Paris will be very different from the one starting in 1990 in Phoenix, implying that the usual scaling form does not apply in general. In this paper, we investigate this question and test if a scaling exponent computed by aggregating data for different cities (usually at a same date) is relevant for predicting what will happen at the level of individual cities as their population grow. We illustrate this discussion on the case of congestion-induced delays but our results could have far-reaching consequences on many other scaling results for cities.

\section*{Aggregating all cities: Global scaling}

\begin{figure}[!h]
\centering
\includegraphics[scale=0.25]{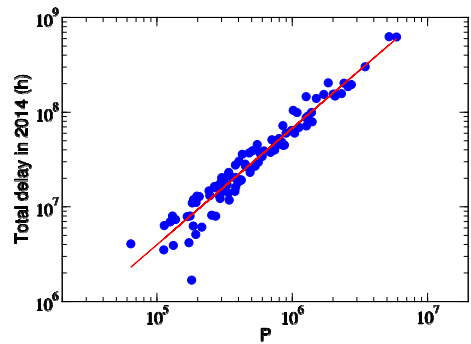}
\caption{Plot of the annual delay $\delta\tau$ versus the number of drivers $P$ for all cities in 2014 (data from TTI's Urban Mobility Information website, see \cite{data}). The straight line is a power law fit in this loglog representation and gives an exponent value $\beta\approx 1.23$ (and $R^2=0.97$).}
\label{fig:2014}
\end{figure}

We focus on the particular case of traffic congestion and its impact on time delays. Previous studies have been made in order to empirically test and theoretically explain how traffic congestion scale with the population.  In \cite{Barthelemy:2016b,Louf:2014} for instance, the authors propose a theory of urban growth which accounts for some of the observed scalings. The theoretical predictions are tested against several data sets, collected by the  Organisation for Economic Co-operation and Development (OECD) or by a GPS device company (TomTom) \cite{Barthelemy:2016b}. Here, we study the dataset (freely available at \cite{data}) published by the Texas A\&M Transportation Institute (TTI) in the Urban Mobility Report (UMR), obtained for $101$ cities in the United States over $33$ years from 1982 to 2014 (the methodology used for constructing this dataset is described in \cite{methodo}, and we also give more details about this dataset in the SI). This database has been investigated in 2017 by \cite{Chang:2017} and in this study, the authors agglomerate all the data corresponding to different cities and performed the usual power law fit of the form
\begin{align}
\delta\tau_i=a  P_i^\beta
\end{align}
where $\delta\tau_i$ is the annual congestion induced delay corresponding to the city $i$.  In this study we take for $P_i$ (also denoted by $P$ in the following) the number of car commuters for the city $i$ rather than the population, because this is the relevant parameter in many models that deal with congestion in cities (see \cite{Louf:2014}). If we take the population instead of the number of car commuters, our results are qualitatively the same and our conclusions remain unchanged, even if all the exponent values change slightly (a fit for all cities and all years shows that the number of car commuters is approximately a constant fraction of order $35\%$ of the population). In \cite{Chang:2017}, they used the least square method to estimate $\beta$ and for the year 2014 (the last available year in the urban mobility report), we find with this method $\beta=1.23 \pm 0.03$. We plot the data and the corresponding fit on Fig.~\ref{fig:2014}.

The quality of a fit has in general to be carefully checked with the help of statistical methods \cite{Leitao:2016}, and computing a good estimation of this exponent values relies on several assumptions: data points are independent, the noise is multiplicative and has a variance independant of $P_i$ (homoscedasticity). It should also be checked that the nonlinear fit that has an additional parameter compared to the linear one, is much better than what would be expected by pure chance.  In this case, the trend seems however to fit the data in a reasonably good way with a large $R^2=0.93$, even if we have only two decades here. The value of $\beta$ larger than $1$ indicates a superlinear behavior of the traffic congestion, a fact in agreement with recent empirical \cite{Chang:2017} and theoretical approaches \cite{Louf:2014,Bettencourt:2013}.

\begin{figure}[!h]
\centering
\includegraphics[scale=0.25]{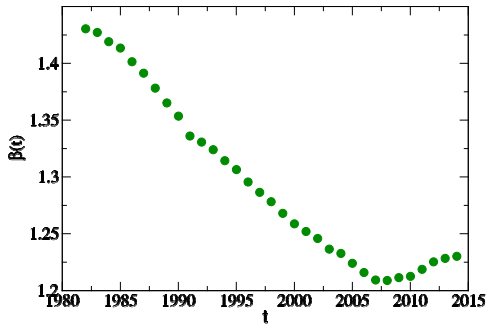}
\caption{Scaling exponent $\beta(t)$ for the delay computed for each year separately, from 1982 to 2014. All these values are consistent with a superlinear behavior found in \cite{Chang:2017}.}
\label{fig:mus}
\end{figure}

\begin{figure}[!h]
\includegraphics[scale=0.25]{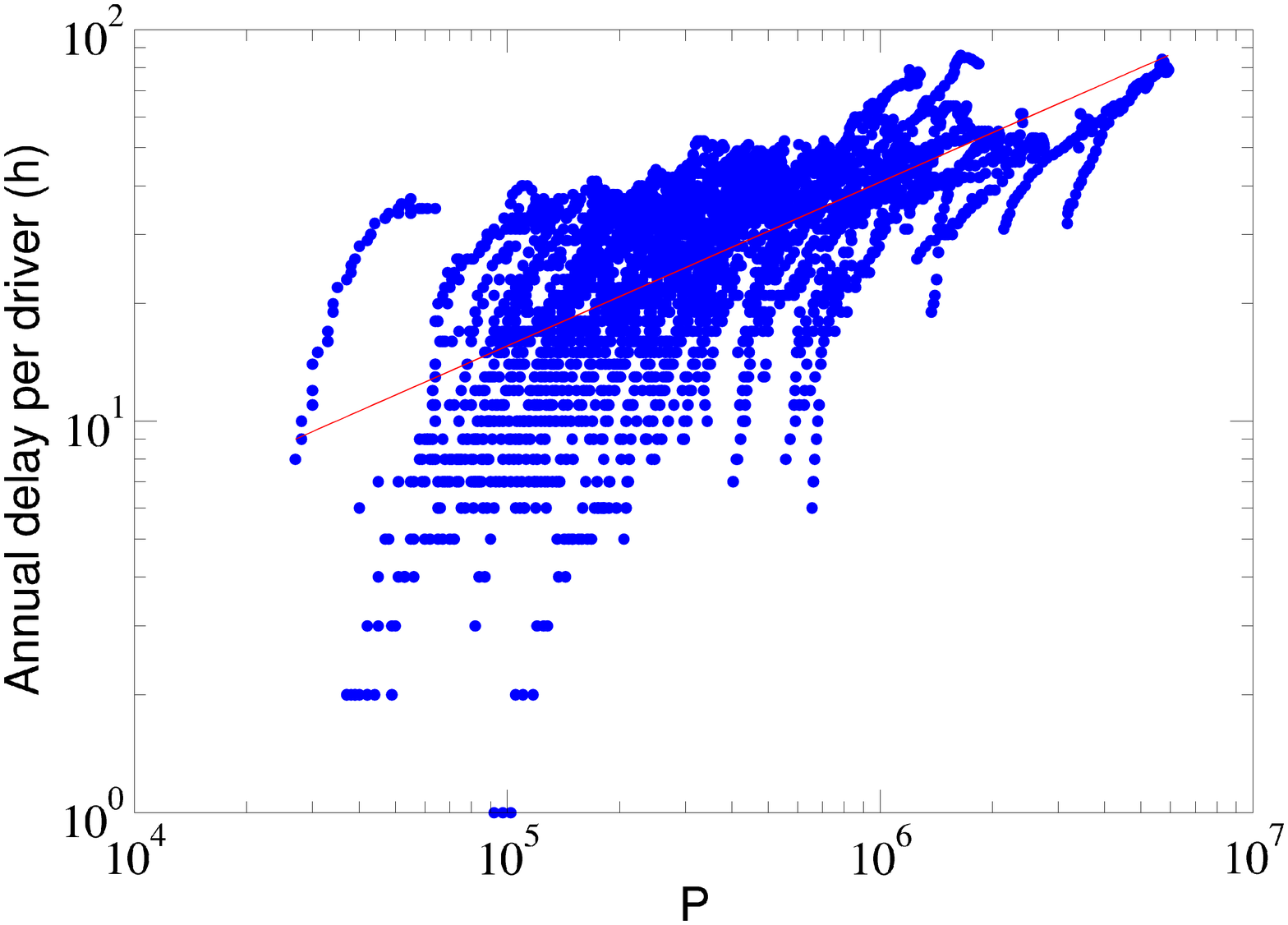}
\includegraphics[scale=0.25]{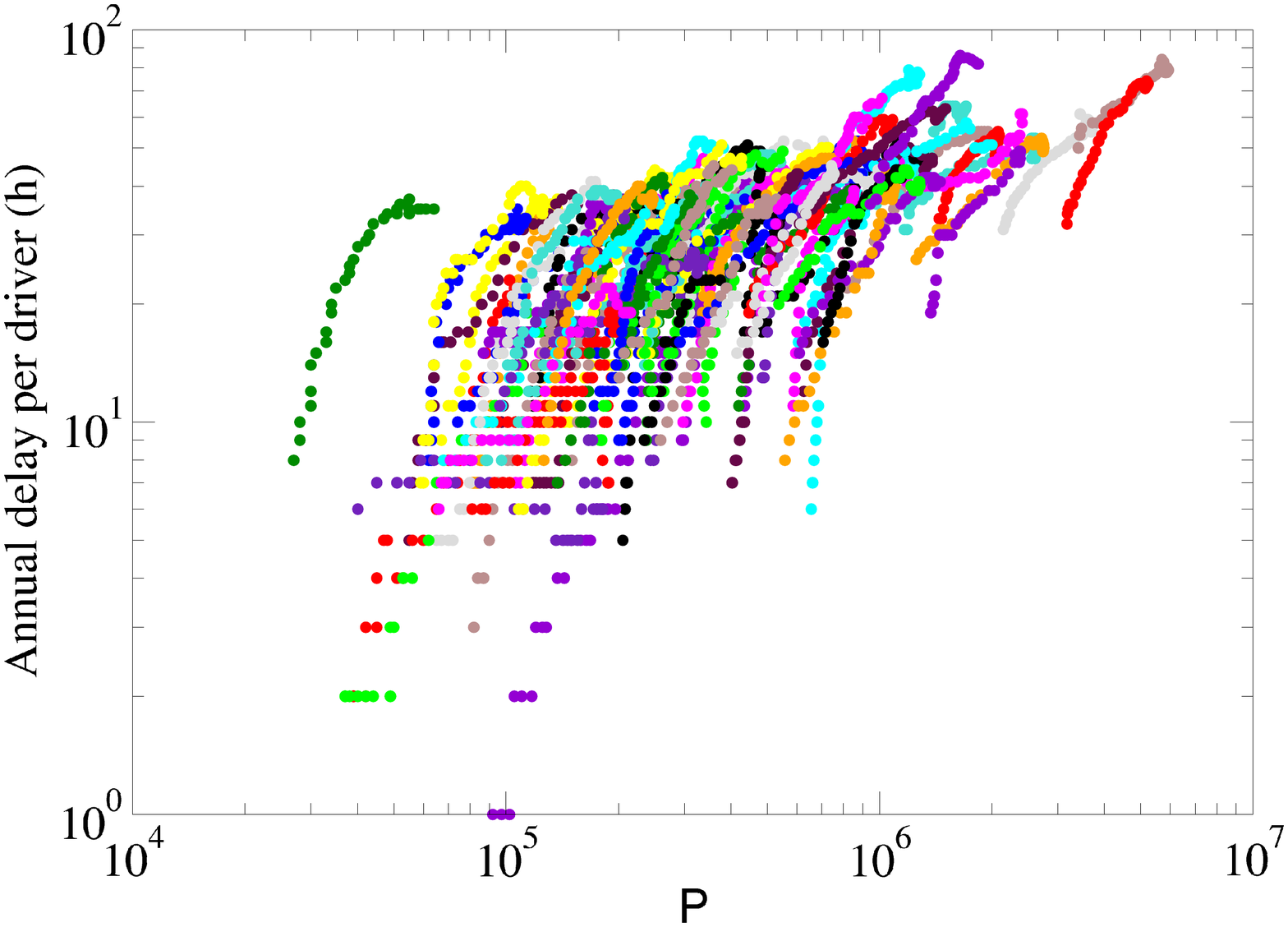}
\caption{(Top) Scatterplot of the annual delay per capita $\delta\tau/P$ versus $P$ for all the 101 cities and for all years (1982-2014). The straight line is the power law fit leading to the value $\beta\approx 1.36$ consistent with a superlinear behavior. (Bottom) Same scatterplot but where the points are colored according to the city they describe (one color per city). As we discuss in the text there is no obvious relation between the global power law scaling and the individual behavior of cities.}
\label{fig:toymodel}
\end{figure}

We can repeat this fit for each year separately, from 1982 to 2014. Formally, we test for each time $t$ the relationship $\log(\delta\tau_i(t))=\log(a)+\beta(t) \times \log(P_i(t))+\text{noise}$, where $\beta(t)$ is the scaling exponent to be determined. We show the values of  $\beta(t)$ versus $t$ in Fig.~\ref{fig:mus} and
we observe that $\beta(t)$ is not constant through time and displays non-negligible fluctuations of order $20\%$. However all these values are larger than $1$ indicating a consistent superlinear behavior. In \cite{Chang:2017} a least square method has been used on all the points available: they mix all the 33 years available for each of the 101 cities and get $33 \times 101= 3333$ points leading to a scaling exponent $\beta\approx 1.36  \pm 0.01$, consistent again with a superlinear relation, as found in \cite{Chang:2017}. For this dataset, we plot the scatterplot and the corresponding nonlinear fit in Fig.~\ref{fig:toymodel}(top) (note that we plot here the delay per capita).
We observe some variability but the global increasing trend seems to be correct. This way of proceeding with data is common: one mixes data for different cities and for the available years, and then performs a regression over the whole set. The scaling that is obtained -- and that we qualify as `global'-- is then used for discussing theoretical approaches. For instance, in \cite{Bettencourt:2013}, this approach is used for computing some scaling exponents (for quantities such as land area, wages, etc.) and are compared with the exponent expected from theoretical calculations. In \cite{Bettencourt:2010}, empirical regularities are found by applying this methodology to different indicators, suggesting the existence of a universal socioeconomic dynamics. Beyond statistical problems related to fitting procedures, the exact meaning and the relevance of this global scaling for individual cities is however not clear. In other words, when we know that a certain quantity $Y$ scales for all cities as $Y\sim P^\beta$, what can we say about the evolution of a single city ? In the following we address this question on the case of congestion delay and by studying in details the dynamics of every individual city and compare its behavior with the global scaling described above.

\section*{The dynamics of individual cities}

In Fig.~\ref{fig:toymodel}(bottom), we show the same plot as in Fig.~\ref{fig:toymodel}(top) but where we now distinguish cities (one color corresponds to one city). This allows us to compare the evolution of the delay due to congestion in each city when its population grows. The first striking observation is that for \emph{all} cities in our dataset, the evolution of the congestion delay \emph{does not behave as predicted by the global trend}. They have their own trend which depends on their particular history. In this respect, it is natural to ask what is the individual city dynamics and what does it have in common with the global scaling. In what follows we thus focus on this individual behavior and discuss its relation with the global power law exponent.

\subsection*{Absence of a single scaling}

\begin{figure}[!h]
\centering
\includegraphics[scale=0.25]{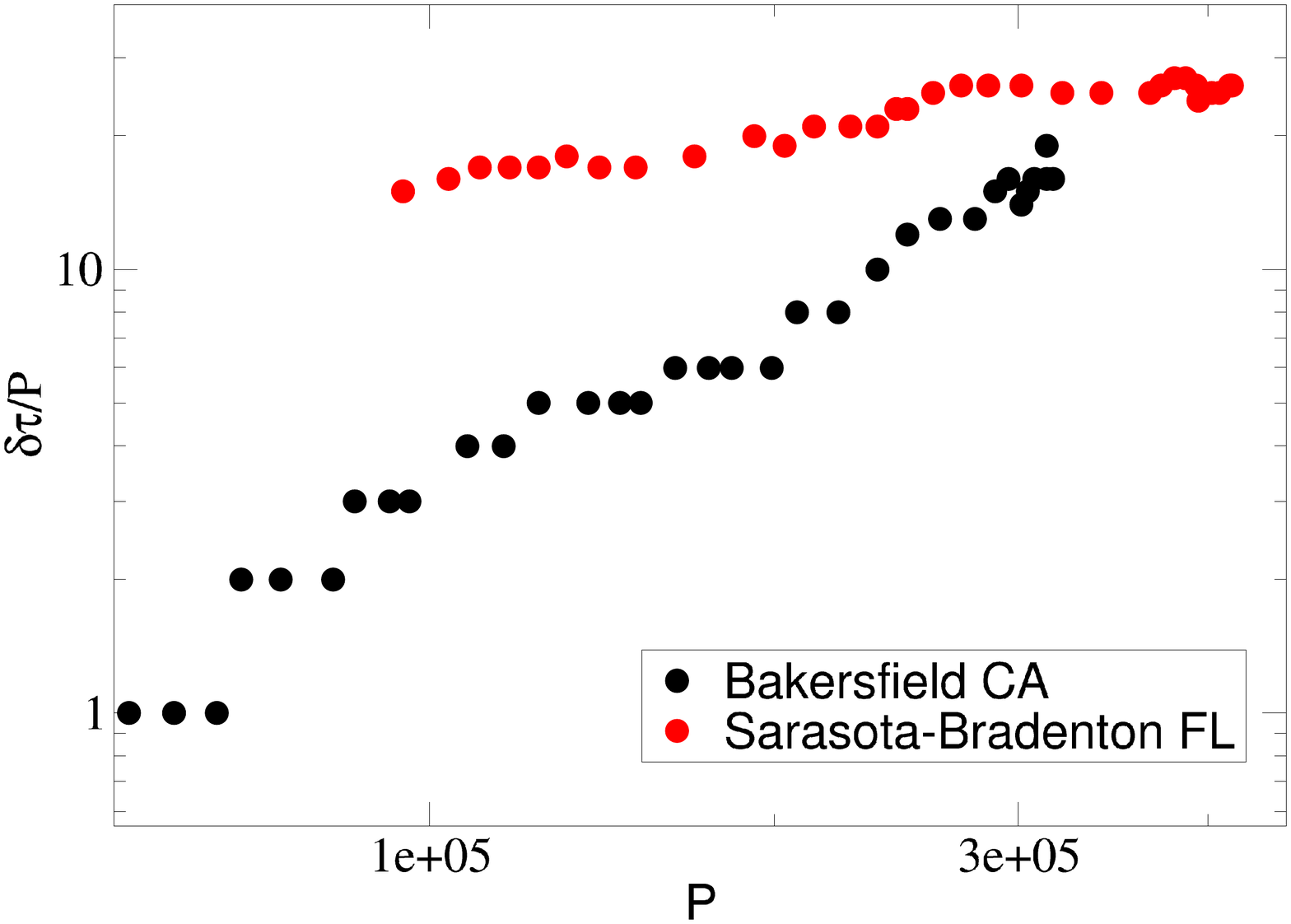}
\includegraphics[scale=0.25]{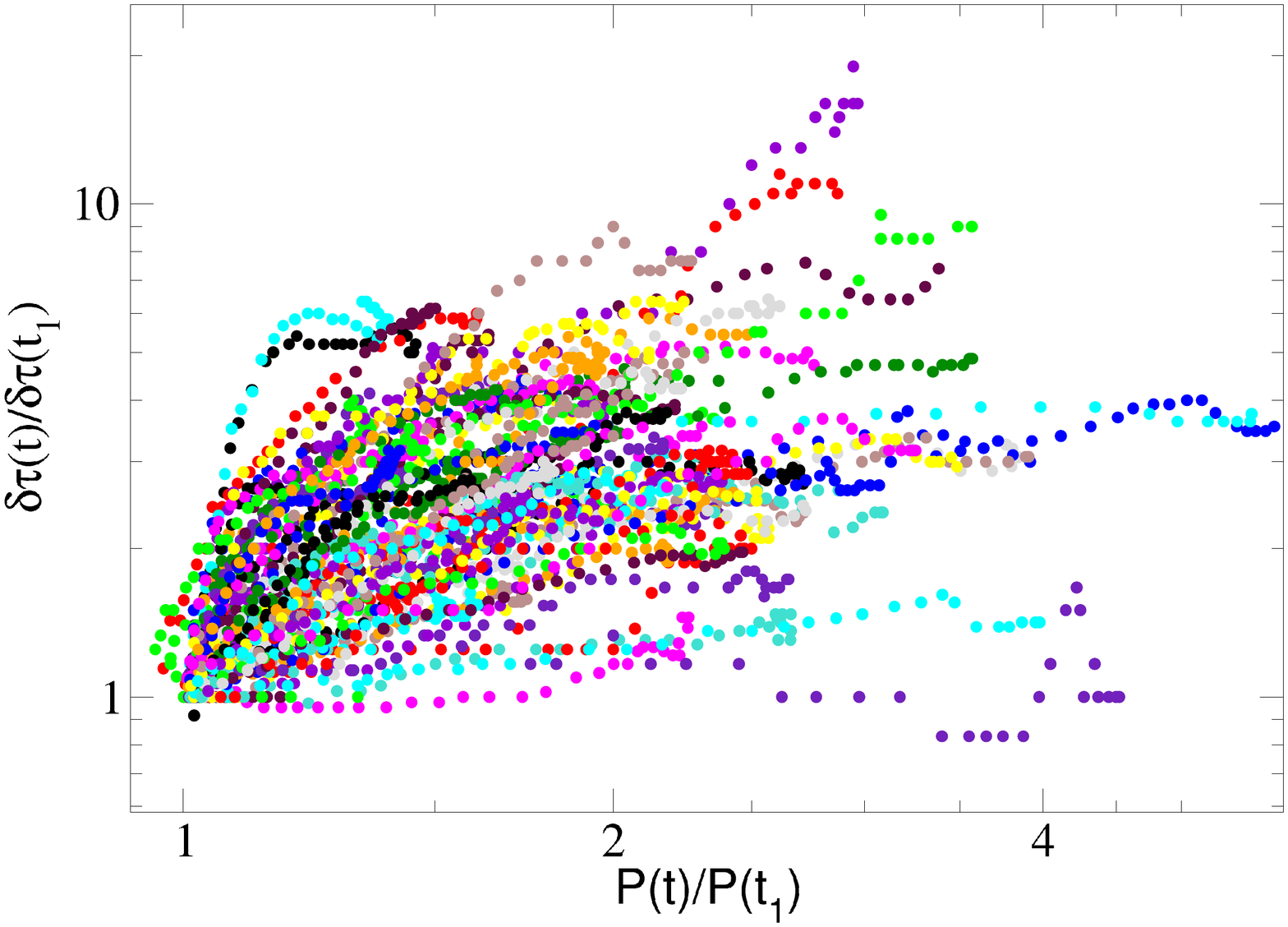}
\caption{(Top) Loglog plot of the annual delay per capita $\delta \tau/P$ versus $P$ for two different cities: Bakersfield (CA) and Sarasota (FL). For the same range of $P$ values, the delay is different, and the slopes are different as well. (Bottom). Plot of the rescaled delay $\delta\tau(t)/\delta\tau(t_1)$ versus $P(t)/P(t_1)$. The curves correspond to different cities and the fact that they do not collapse indicates the absence of a unique scaling determined by a single exponent.}
\label{fig:deuxvilles}
\end{figure}

With this dataset, we can monitor the evolution of each city when its population grows. The first thing that
we observe on the examples in Fig.~\ref{fig:deuxvilles}(top) is that the annual delay is not a simple function of $P$ only. The value of the number of drivers (or the population) is not enough to determine the delay. We also note in this figure that the slopes are different (a power law fit gives $\beta\approx 3.20$ for Bakersfield and $\beta\approx 1.45$ for Sarasota) showing that even when a power law exists it is not with the same exponent (see section `Type-1 cities' below for a further analysis of this point). In order to test further the existence of a scaling of the form $\delta\tau\sim P^\beta$ we 
plot in Fig.~\ref{fig:deuxvilles}(Bottom) for all cities $\delta\tau(t)/\delta\tau(t_1)$ versus $P(t)/P(t_1)$ where $t_1$ is the first available time.
Even if the prefactor changes from a city to another this rescaling allows to test the existence of a unique power law scaling. As we can see in this figure~\ref{fig:deuxvilles}(bottom), the curves for different cities do not collapse signalling the absence of a scaling form governed by a single exponent. In the following we will focus on the different behaviors observed for this set of cities. 

\subsection*{Different categories of cities}

\begin{figure}[!h]
\centering
\includegraphics[scale=0.25]{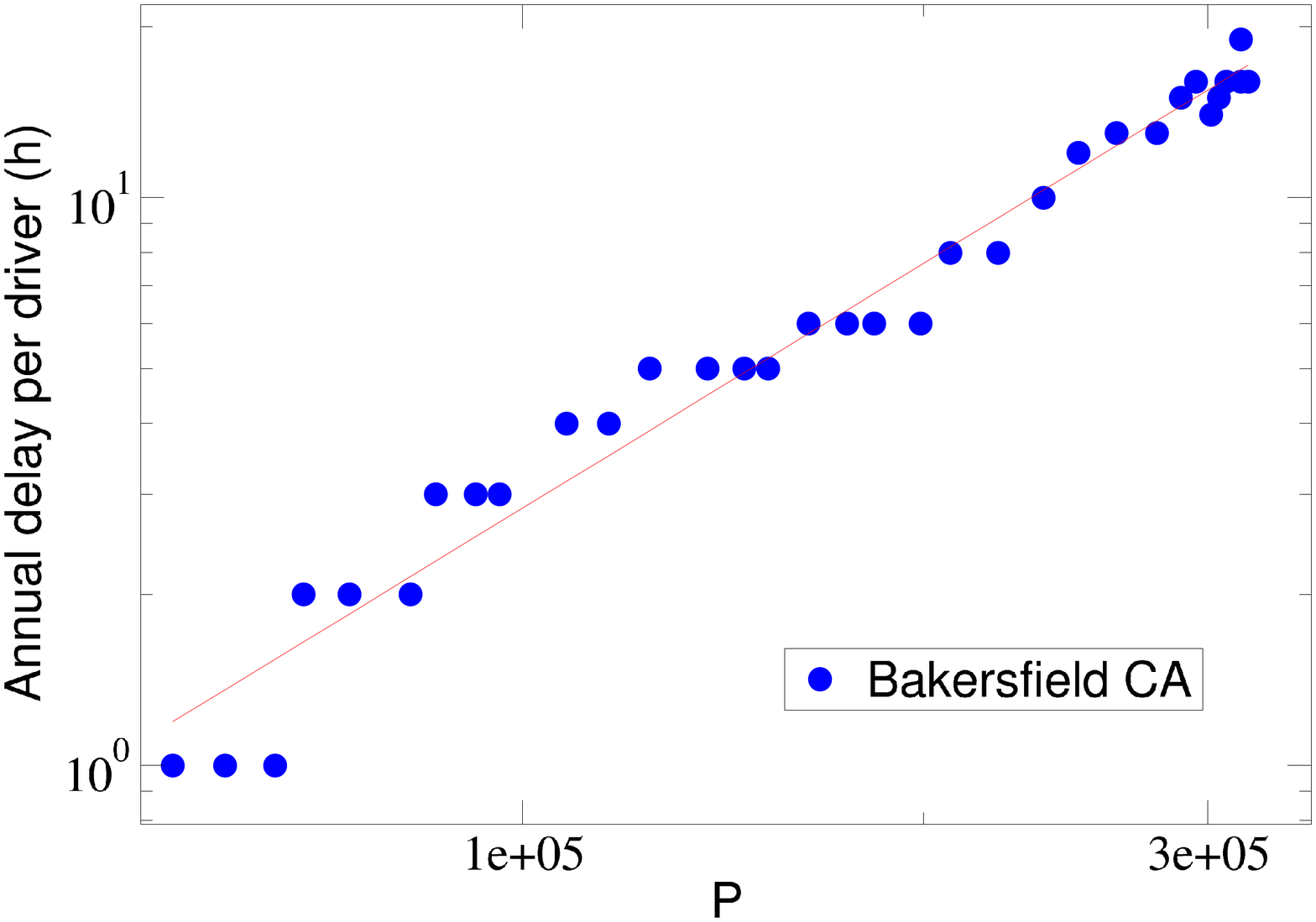}
\includegraphics[scale=0.25]{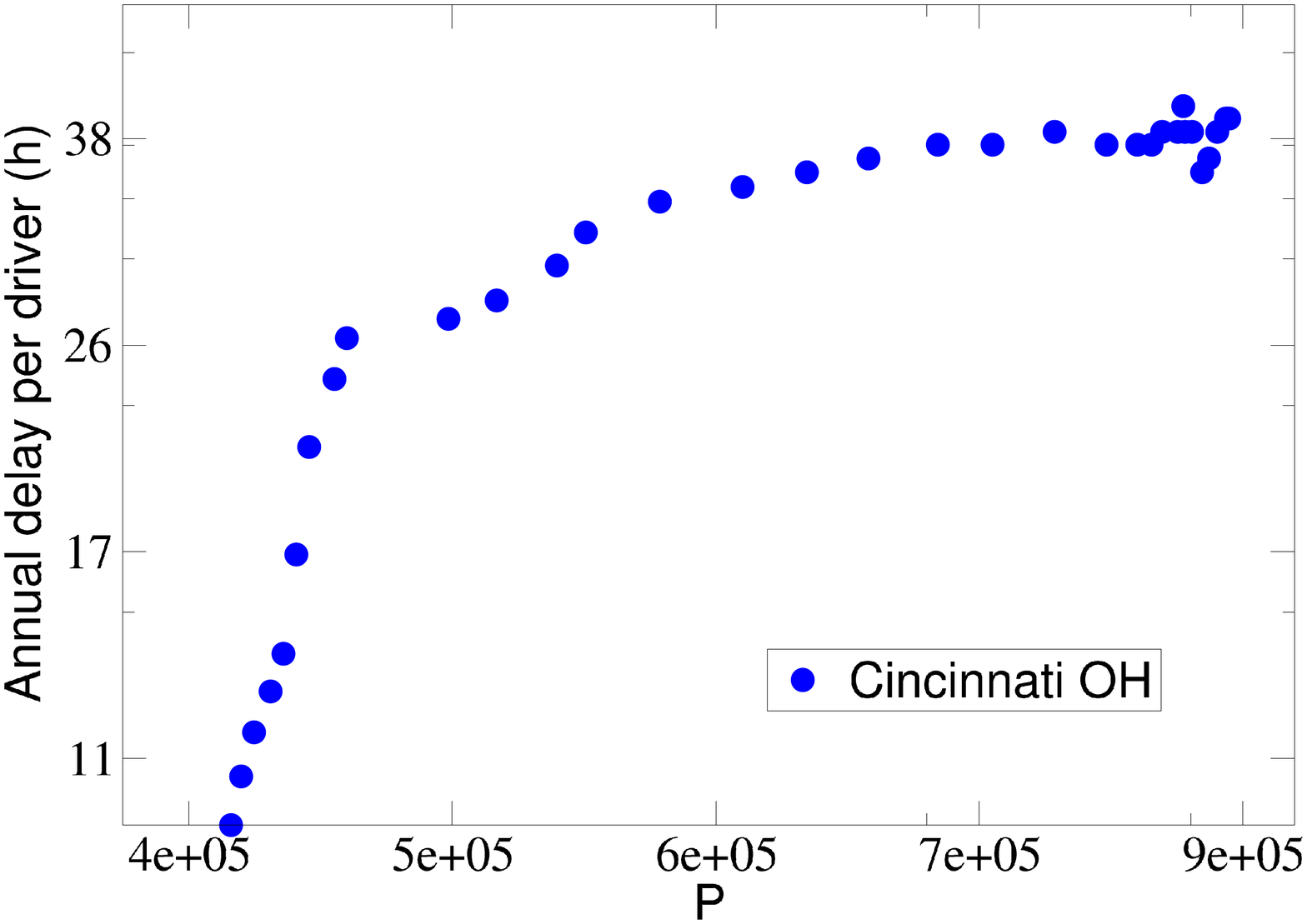}
\caption{Loglog plot of the annual delay per capita $\delta \tau/P$ versus $P$ from 1982 to 2014. (Top) An example of a type-1 city where the delay grows with $P$ and that can be reasonably fitted by a power law (Bakersfield, CA). (Bottom). Example of a type-2 city with two power law regimes characterized by two different exponents (Cincinatti, OH).}
\label{fig:uneville}
\end{figure}

We analyze the behavior of each of the 101 cities in the dataset and we observe a variety of behaviors. 
More precisely, there are two main categories characterized by different time evolutions:
\begin{itemize}
\item The delay increases with $P$ and in most cases can be fitted by a power law (see Fig.~\ref{fig:uneville}(top)) and we refer to this set as `type-1' cities and which represent over $30\%$ of our cases. We note here that for the dataset studied here, the time range (from 1982 to 2014) does not allow to have a very large variation of the number of drivers (the ratio $P(2014)/P(1982)$ varies from $1.2$ to $6$ approximately) and a much larger dataset would be needed in order to have a better accuracy for these exponent values. 

\item The other cities (about $40\%$ of all cities) display two regimes separated by a change of slope that is in general abrupt. The second regime for these `type-2' cities can be in some cases a `saturation' where the delay stays constant. We show in Fig. ~\ref{fig:uneville}(bottom) an example of such city that displays saturation with a zero slope in the second regime.  

\item The rest of cities ($\approx 30\%$) do not display a common behavior (for instance some present two or three changes of slope, etc.)
\end{itemize}
In most cases however, the individual behavior of a city does not correspond to the global scaling $\delta\tau\sim P^{1.36}$. 
In the following we focus on each of these classes and try to characterize them more precisely. 

\subsection*{Type-1 cities: power law growth}

\begin{figure}[!h] 
\centering 
\includegraphics[scale=0.25]{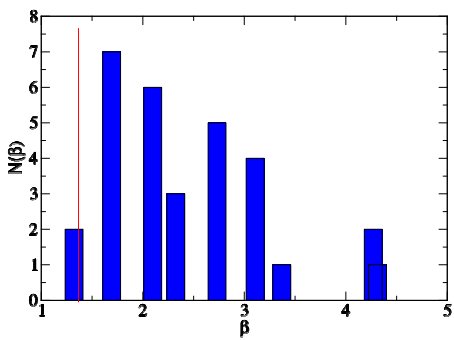} 
\caption{Empirical histogram of $\beta$ for type-1 cities. The average value is $2.46$ (and the dispersion is $\sigma=0.91$). The vertical line indicates the value of the global scaling $\beta\approx 1.36$.}  
\label{fig:histtype1} 
\end{figure}

This particular class comprises cities that display an individual scaling law that can be fitted by a power law of the form $\delta\tau(t) \simeq P(t)^{\beta(i)}$ , where $P(t)$ is the number of commuters at time $t$ and $\delta\tau(t)$ the corresponding annual congestion-induced delay. The quantity $\beta(i)$ depends in general on the city $i$ and we show in Fig.~\ref{fig:histtype1} the histogram for this exponent computed for all type-1 cities.  We clearly see that very few cities behave as the `global trend' predicted: only 2 cities over 31 have an exponent $<1.5$, while 13 cities have an exponent $>2.5$ (we give in the SI Appendix, the list of values for $\beta$). This result shows that when we observe a power law behavior at the individual city level, it is generally with an exponent that is much larger than 1 and much larger than the result found for the global scaling. In other words there seems to be no correlation between the global observation made on all cities and the individual behavior of cities when its population evolves.

\subsection*{Type-2 cities: existence of two regimes}

\begin{figure}[!h]
\centering
\includegraphics[scale=0.25]{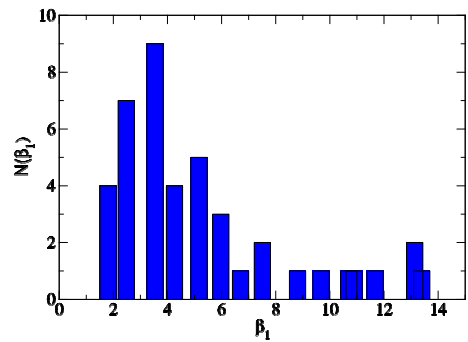}
\includegraphics[scale=0.25]{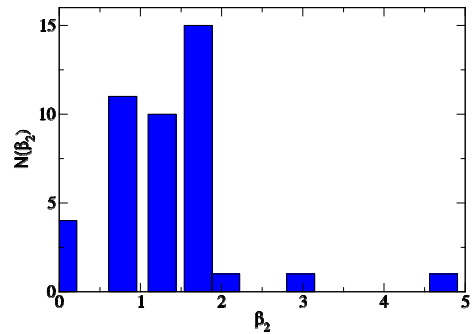}
\caption{Empirical histograms for the two exponents $\beta_1$ and $\beta_2$ that describe the two regimes of type-2 cities. (Top) Histogram for $\beta_1$. The average is $\overline{\beta}_1\approx 5.35$ and the dispersion is $\sigma\approx 3.31$. (Bottom) We show the histogram for $\beta_2$. The average is here $\overline{\beta}_2\approx 1.32$ with a dispersion $\sigma\approx 0.81$. For most cities we have $\beta_1>\beta_2$.}
\label{fig:histtype2}
\end{figure}

\begin{figure}[!h] 
\centering 
\includegraphics[scale=0.25]{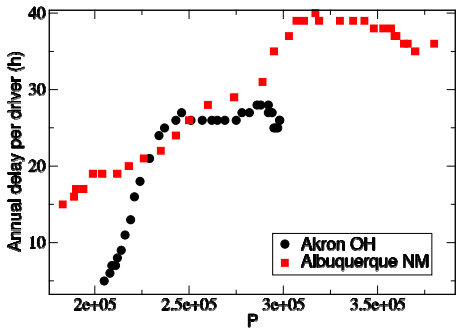} 
\caption{Example of two different type-2 cities with two regimes characterized by two exponents $\beta_1$ and $\beta_2$. In the case of Akron (OH) we observe a `saturation' with a constant delay per capita ($\beta_2\approx 1$), while for Albuquerque (NM) the delay per capita decreases with the population ($\beta_2<1$).}
\label{fig:Akron}
\end{figure}

For about $40\%$ of the cities in the dataset, the delay versus the number of car commuters displays a change of slope and $\log(\delta \tau)$ is a piecewise linear function of $\log(P)$. Formally one could write:
\begin{align}
\log(\delta \tau) = 
\begin{cases}
a_1+\beta_1\times \log(P) \text{ when } P< P^* \\
a_2+\beta_2\times \log(P) \text{ when } P>P^*
\end{cases}
\end{align}
This behavior indicates that the dynamics of the traffic congestion in those cities followed successively two different scaling laws with two different exponents $\beta_1$ and $\beta_2$ and we plot the histograms for both these exponents in Fig.~\ref{fig:histtype2} (we give in the SI Appendix, the list of values of $\beta_1$ and $\beta_2$).
We note that the average of $\beta_1$ is around $5.35$, while the average of $\beta_2$ drops to $1.32$, closer to the `global exponent' (but with a large dispersion around this value). Beyond averages, we have that for almost every case, $\beta_1> \beta_2$ (we also show in the SI Appendix that there are no correlations between $\beta_1$ and $\beta_2$). Almost all the exponents of the first regime $\beta_1$ are above 2 (indicating a strong superlinearity) while the second exponents $\beta_2$ are mostly $< 2$. For this second regime, some cities do not exhibit superlinear behaviour. Indeed for some cities ($\sim 30 \%$), the exponent $\beta_2$ is very close to $1$, indicating a linear behavior and equivalently a delay per capita constant -- that we coined `saturation'. The cities of Akron (see Fig.~\ref{fig:Akron}), or Birmingham for instance fall into that subcategory.  
We also observe that in some cases a crossing between the curves corresponding to different cities can occur (such as Akron and Albuquerque in Fig.~\ref{fig:Akron}). This crossing is another sign that the posterior evolution of a city is not uniquely determined by the population and the delay at a certain time (if it did the evolution after the crossing should be identical for the two cities).

\begin{figure}[!ht]
\centering
\includegraphics[scale=0.25]{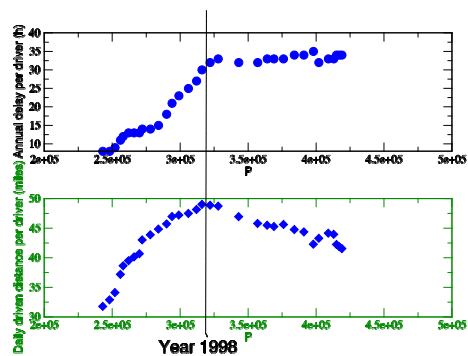}
\caption{Birmingham case. (Top) Loglog plot of  $\delta \tau/P$ versus $P$. (Bottom) Loglog plot of the total driven length per capita $L_{tot}/P$ vs $P$. The vertical dotted line indicates the change of slope of  $\delta\tau/P$ and corresponds here to the year 1998.}
\label{fig:LtotBirmingham}
\end{figure}

In other cases ($\sim 10\%$), the exponent $ \beta_2$ is clearly $<1$, which indicates sublinearity and that the delay per capita decreases with the population. We show the example of the city of Albuquerque (New Mexico) in Fig.~\ref{fig:Akron}. This phenomenon is very counter intuitive, even if we can point out some elements of explanation. Indeed, in addition to the congestion induced delay, we also have the data for the total driven length $L_{tot}$ (in $miles\times commuters$) for each city and each year. We can check if this quantity can explain, even partially, the behavior of the total delay. For some type-2 cities with two regimes, we plot the driven length per commuter against the number of drivers and we observe that this curve displays a change of regime at the exact same point for the delay. In Fig.~\ref{fig:LtotBirmingham}(top), we see that for the case of Birmingham, from 1998, the delay remains almost constant, whereas it increased constantly at a high rate before that (more precisely we have here $\beta_1\simeq 5.7$ and $\beta_2 \simeq 1$). In Fig.~\ref{fig:LtotBirmingham} (bottom), we observe that in the same year, the curve for $L_{tot}/P$ experienced a change of slope: the length per capita increased before 1998, and slowly decreases after that date. This could explain partially why the delay does not evolve after this date: there are certainly more people on the road after 1998, and therefore more likely some congestion, but each commuter drives less on average which decreases the occurrence of traffic jams: these two effects can compensate each other. This is one possible partial explanation, which however does not hold for all the cities. The change of slope in $L_{tot}/P$ vs $P$ is common in this dataset and in most cases happens simultaneously with the change of regimes of the delay, pointing to the existence of correlations between these quantities, even if not in a causal manner. The simultaneous change of regime for these two quantities might also be the sign that the city experienced a large scale structural change. 

For this category of cities, beyond the two exponents $\beta_1$ and $\beta_2$, we can also study (i) at what \emph{time} $T^*$ the change of slope happened, (ii) what was the \emph{population} of the city when it happened ($P^*$), and (iii) what was the delay par capita when it happened ($(\delta \tau/P)^*$). The histograms for these quantities are shown in the SI Appendix, Fig. S5. The distribution of $T^*$ is difficult to interpret and does not display a typical date at which the slope changes. The change of slopes therefore does not occur at the same time for these cities, which would have been the case for instance if there had been a national plan in the US to rebuild the whole road system, or any other federal decision. The histogram for $P^*$ seems clearer to interpret with the existence of a clear maximum around $200,000$ commuters and a quick decay for larger values. The average of the distribution is $394,000$, while the standard deviation is $367,000$. Finally, the delay per capita $(\delta \tau/P)^*$ displays a histogram that has a relatively small compact support, with an average of about $39$ hours per year, and a standard deviation about $18$ hours per year. 
This relatively small variation of $(\delta \tau/P)^*$ suggests that it is the congestion that triggers the change of regime signalled by different exponents. Further studies are however certainly needed in order to clarify this important point. 

\section*{Discussion}

We focused in this paper on the dataset for congestion-induced delay in some US cities. This is a particularly interesting dataset as it is both transversal (it contains many cities), and longitudinal (for each city we have the temporal evolution of the delay). This is a rather rare case at the moment, but this type of data will certainly become more abundant in the future and will allow to test our results on other quantities. Our observations about scaling might therefore have far reaching consequences for the quantitative study of urban systems, well beyond the case of congestion induced delays.

The general scaling form $Y\sim P^\beta$ indicates that if the population is multiplied by a factor $\lambda$ the quantity $Y$ is then multiplied by a factor $\lambda^\beta$. This scaling form relies however on a strong implicit assumption which is the `logarithmic population translation' invariance. In other words, this scaling form implies that for any times $t$ and $t'$ we have $Y(t')/Y(t)=(P(t')/P(t))^\beta$ and then depends on the ratio of populations only (or the difference of logarithms). As we observed in this study, there is no such scaling at the individual city level but a variety of behaviors. In the language of statistical physics, the quantity $Y$ (here equal to $\delta\tau$) is not a state function determined by the population only, and displays some sort of aging effect where the delay in a city depends not only on the population but also on the time, and probably on the whole history of the city. In any case we cannot make for a given city a prediction for time $t_2 > t_1$ knowing only its state for $t_1$.  This idea of path-dependency is natural for many complex systems, and in statistical physics, we know that spin-glasses \cite{Bouchaud:1997} for example display aging which means that some features of the system (for instance the relaxation time) evolves with the age of the system and does not depend on the state of the system only. This in particular implies that we do not have time translation invariance but that most functions of two times $t$ and $t'$ do not depend on $t-t'$ only. This aging theory has been applied to many other complex systems, from `soft material' \cite{Fielding:2000} to superparamagnet \cite{Sasaki:2005}, and it would be interesting to understand it in the framework of the evolution of urban systems. An interesting direction for future research would be to investigate the relation between the growth rate of a city and the importance of aging. We could for example test the naive expectation that a slow enough `adiabatic' growth would imply that the size of the city is very important, while a rapide growth could imply that the state of the system at previous times becomes relevant.

The results presented in this paper illustrated on the case of congestion-induced delays could in principle be applied to any other quantity. They highlight the risk of agglomerating data for different cities and to consider that cities are scaled-up versions of each other (as questioned in \cite{Thisse:2014} for example): there are strong constraints for being allowed to do that such as path-independence, which is apparently not satisfied in the case of congestion, and which should be checked in each case.

Beyond scaling, these results also pose the challenging problem of using transversal data (ie. for different cities) in order to get some information about the longitudinal series for individual cities. This is a fundamental problem that needs to be clarified when looking for generic properties of cities.

\paragraph*{Acknowledgments.}
JD thanks the ENSAE and the IPhT for its hospitality. We also thank both
anonymous referees for helpful and interesting suggestions.

\section*{Bibliography}


\bibliographystyle{prsty}

\clearpage
\onecolumngrid
\section{Supplementary Information for `From global scaling to the dynamics of individual cities'}

\subsection{Dataset description}

The dataset is freely available \cite{data} and the methodology is described in the Urban mobility report, 2012 of
the  Texas A\&M Transportation Institute (TTI), College
Station, Texas \cite{methodo}.

This dataset has also been studied in \cite{Chang:2017} and contains the total
hours of delays, excess fuel consumption, and excess $CO_2$ emission
due to congestion for $101$ of the largest urban centers in the
US. The data spans a 30-year period from 1982 to 2011. Other
information such as the population size, number of commuters, the
freeway's lane-miles, and the lane-miles of arterial streets, are also
available at the same source.

\subsubsection{Population size}

\begin{figure}[!h]
\centering
\includegraphics[scale=0.40]{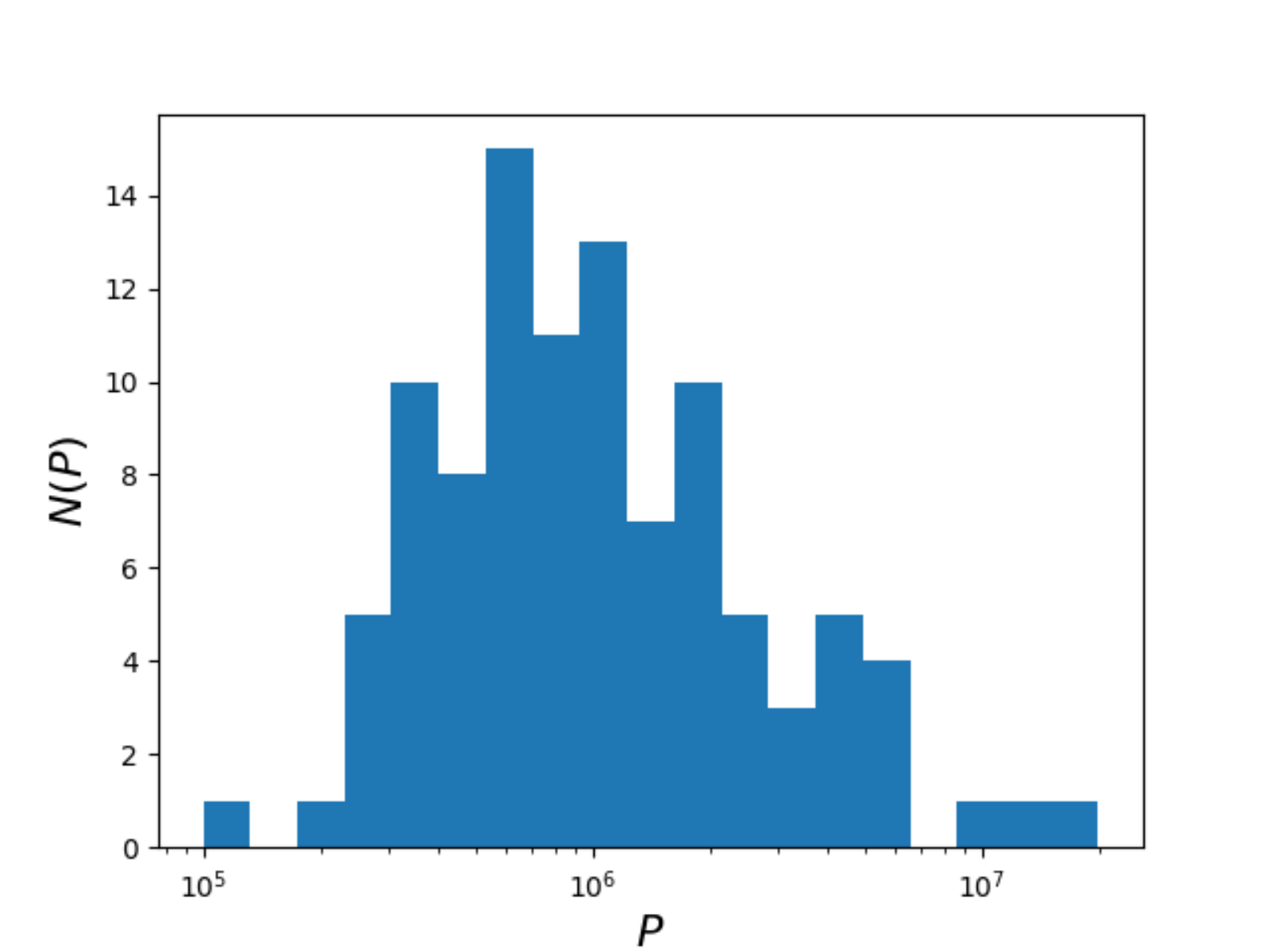}
\caption{Histogram for the population of the 101 cities considered in
  the dataset used in this study.}
\label{fig:histoP}
\end{figure}

The group of the $101$ urban centers described in this dataset
is very heterogeneous and contains cities with very different
population (see Fig.~\ref{fig:histoP}).
We see on this figure that indeed the population of the $101$ cities varies
from $10^5$ to very large numbers of the order $10^7$. 

\subsubsection{Spatial distribution of cities}

\begin{figure}[!h]
\centering
\includegraphics[scale=0.25]{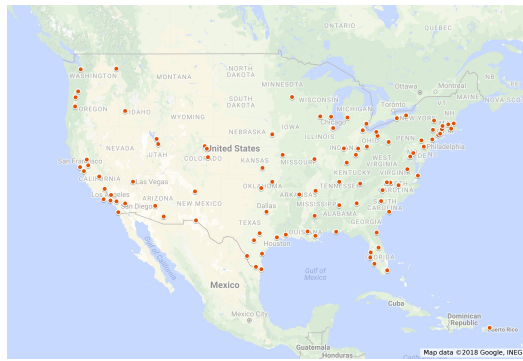}
\caption{Spatial distribution of the cities studied in the dataset [19].}
\label{fig:map}
\end{figure}

The spatial distribution of the cities in this dataset appears to be uniform as can be
seen on the map shown in Fig.~\ref{fig:map}.
This points to the probable absence of spatial bias in the selection of these cities.

\subsection{Exponents}

\subsubsection{Type-1 cities}

For this set of cities, the total annual delay behaves as
\begin{align}
\delta\tau\sim P^\beta
\end{align}
We report in the table \ref{tab:tab1} the list of values for the
exponent $\beta$ for cities in this set.

\begin{figure}[!h]
\centering
\includegraphics[scale=0.40]{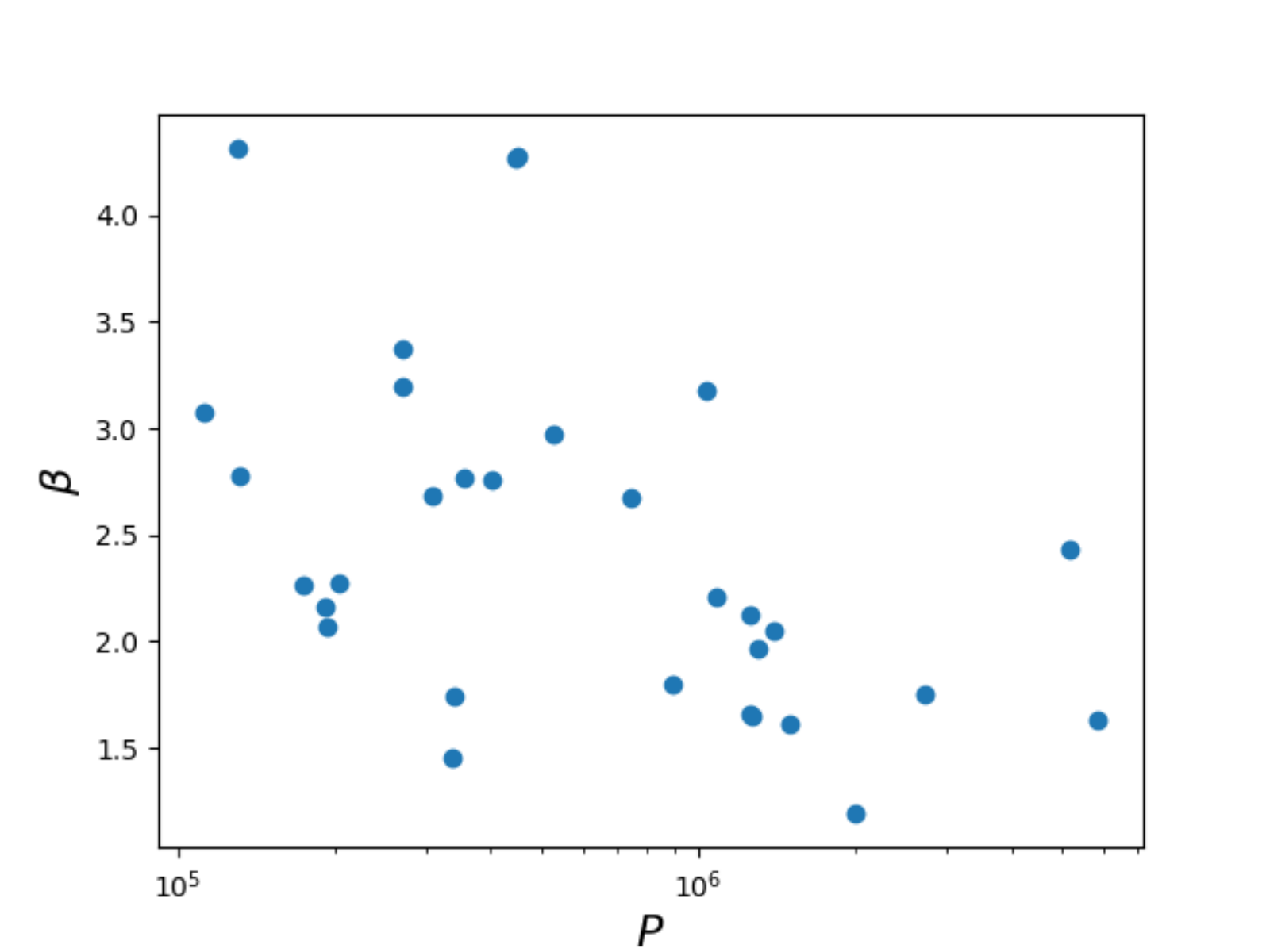}
\includegraphics[scale=0.40]{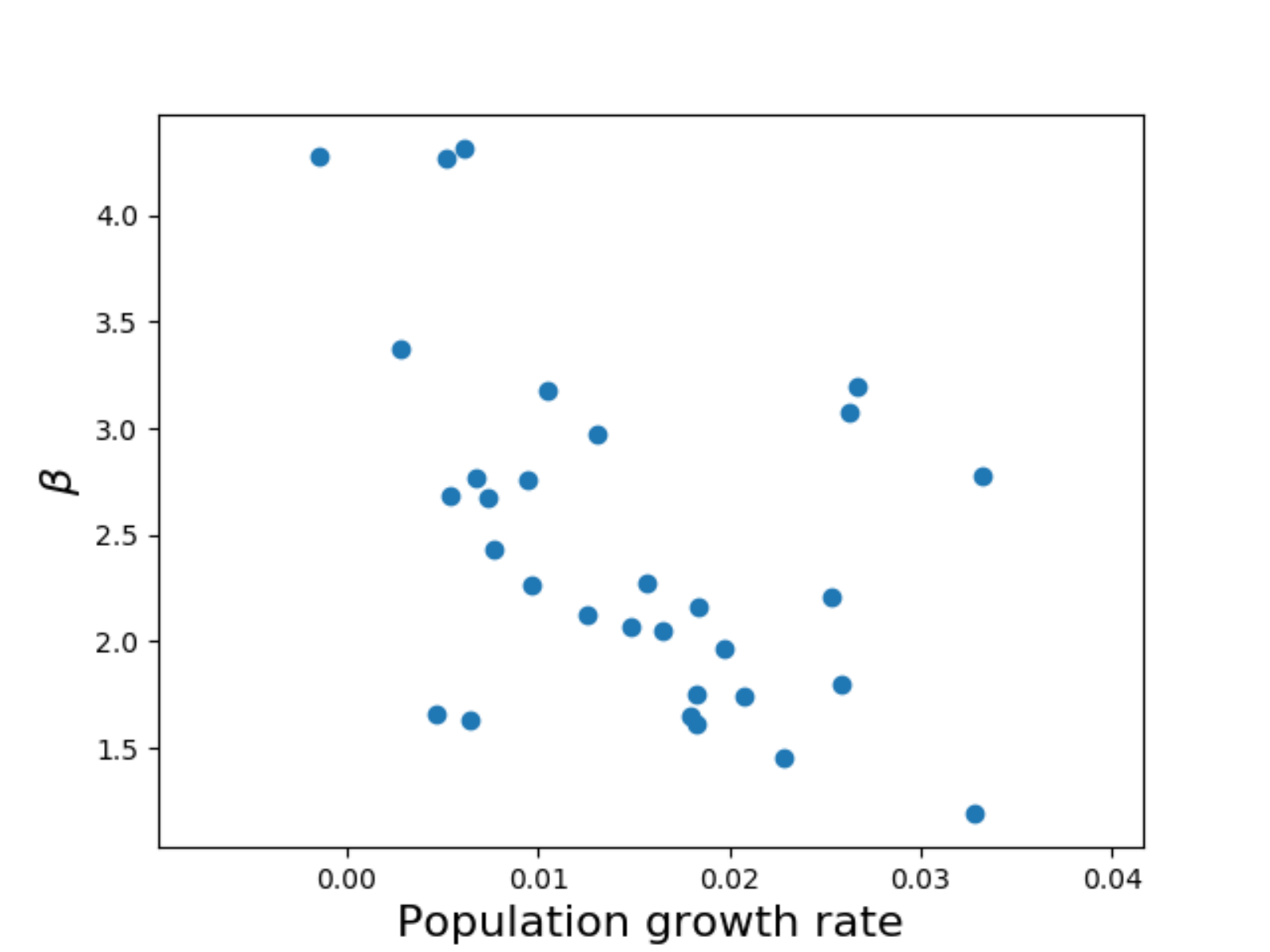}
\caption{Exponent $\beta$ for type-1 cities versus (left) the population in
  2014 and (right) the average population growth rate.}
\label{fig:betavsP}
\end{figure}

We checked that the exponent is not correlated with for example the
final value of the population (see Fig.~\ref{fig:betavsP}, left), but
seems to display some non-negligible correlation with the average
growth rate of a city (Fig.~\ref{fig:betavsP}, right): a linear fit
gives a value of $-42.8$ and a $p-$value of order $1\%$ (we have however to be
careful with these results as the number of points is small
$35$). Certainly more work is needed here in order to study and
understand these correlations.

\subsubsection{Type-2 cities}

By definition, for these cities, the total annual delay displays two regimes: 
\begin{align}
\delta \tau\sim 
\begin{cases}
P^{\beta_1}\text{ when } P< P^* \\
P^{\beta_2} \text{ when } P>P^*
\end{cases}
\end{align}
We report in the table \ref{tab:tab2} the values for the exponents
$\beta_1$ and $\beta_2$ computed for cities in this set.

\subsection{Correlation between $\beta_1$ and $\beta_2$}

\begin{figure}[!h]
\centering
\includegraphics[scale=0.40]{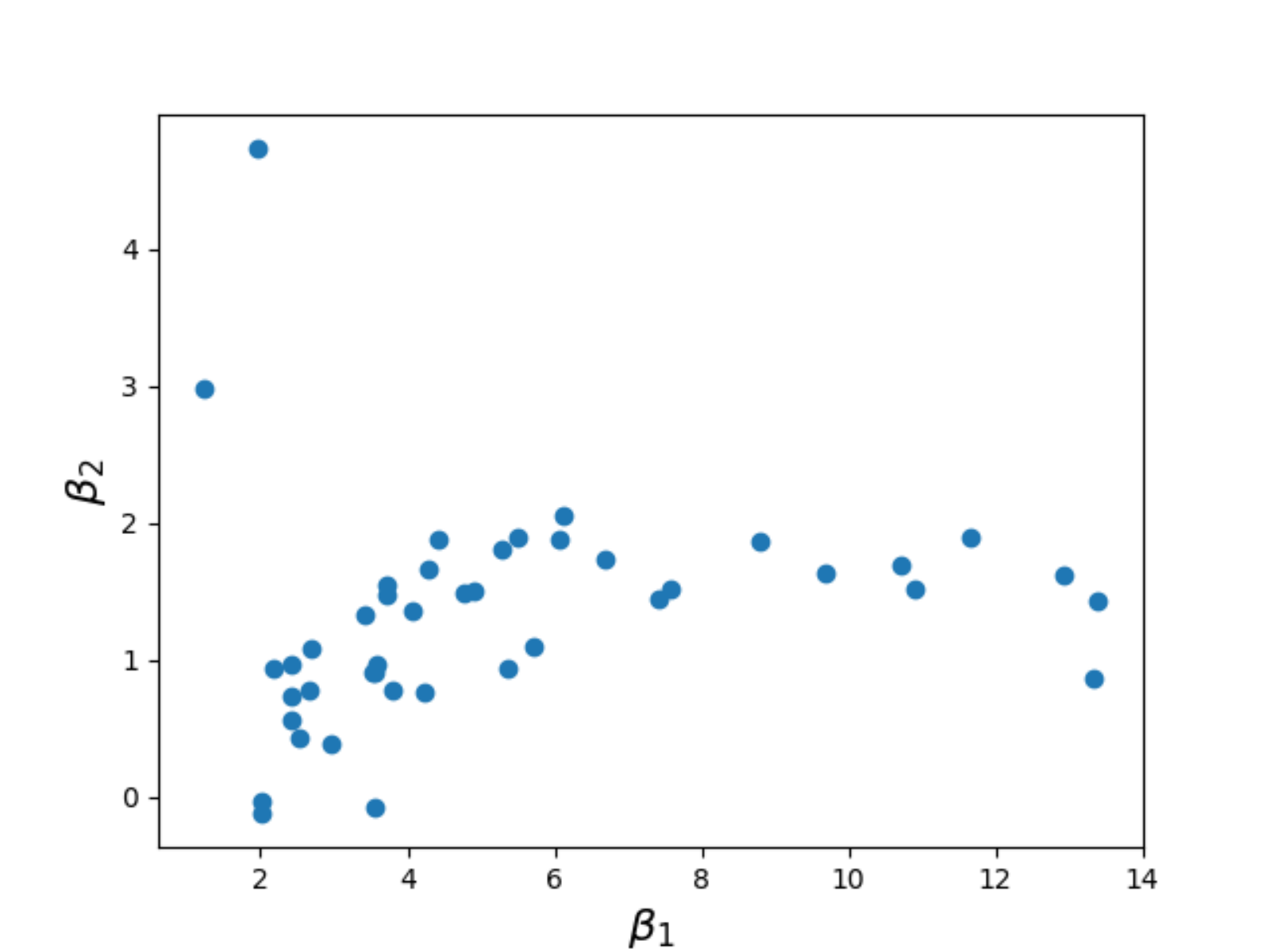}
\caption{Type-2 cities: exponent $\beta_2$ versus $\beta_1$.}
\label{fig:corr}
\end{figure}

For type-2 cities we plot $\beta_2$ versus $\beta_1$ in the
Fig.~\ref{fig:corr}.
We observe in this figure that there are no
significant correlations between these exponents (the $p-$value for the
regression is $0.27> 0.05$ and we cannot reject the hypothesis of no
correlations).

\subsection{Distribution of $T^*$, $P^*$, $(\delta\tau/P)^*$}

For type-2 cities, we show here the distributions
of the quantities defined at the change of slope: $T^*$ is the time at
which the slope happened, $P^*$ is the corresponding population and
$(\delta\tau/P)^*$ is the delay per capita when it happened. 
\begin{figure}[!h] 
\includegraphics[scale=0.40]{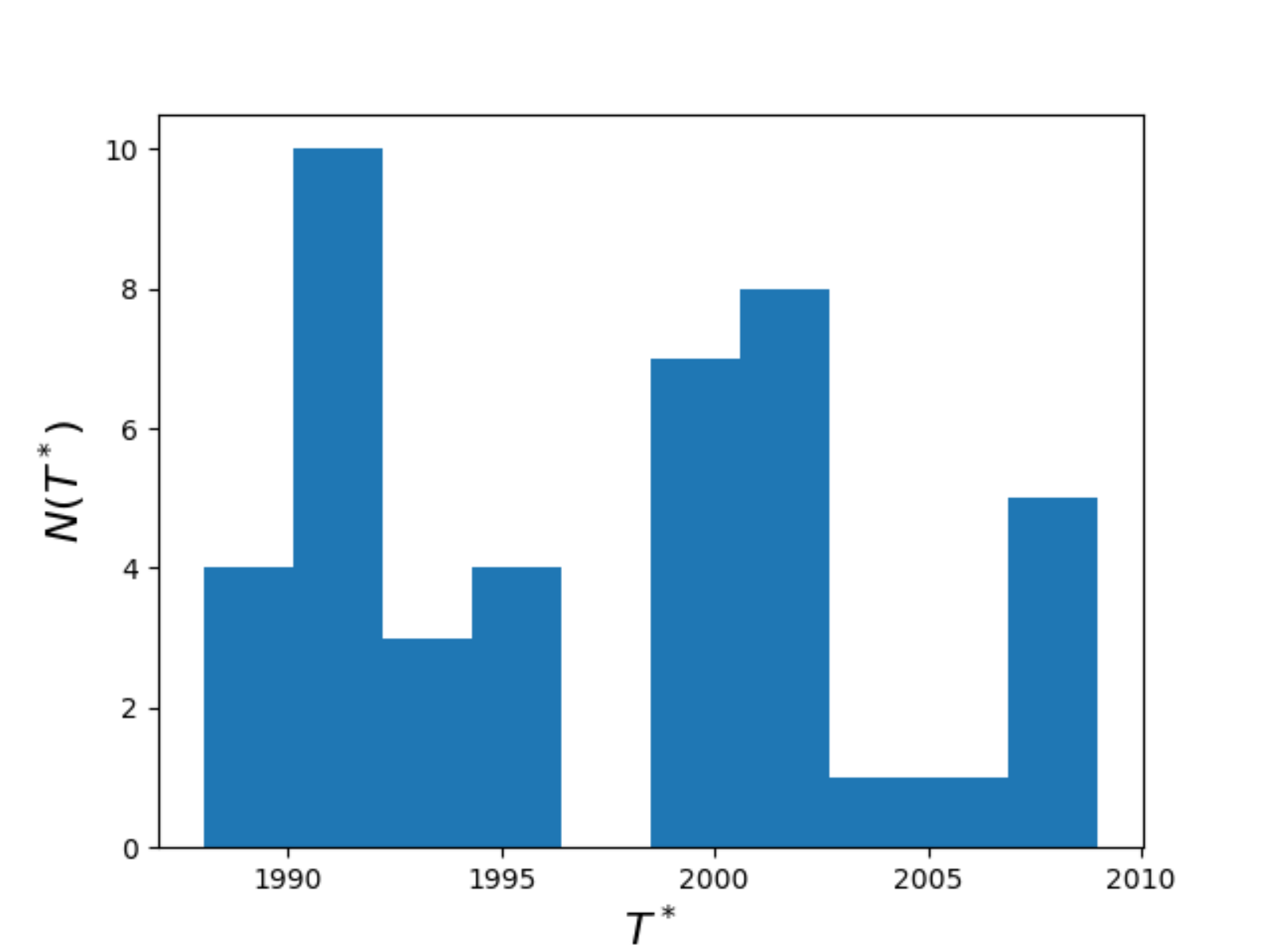} 
\includegraphics[scale=0.40]{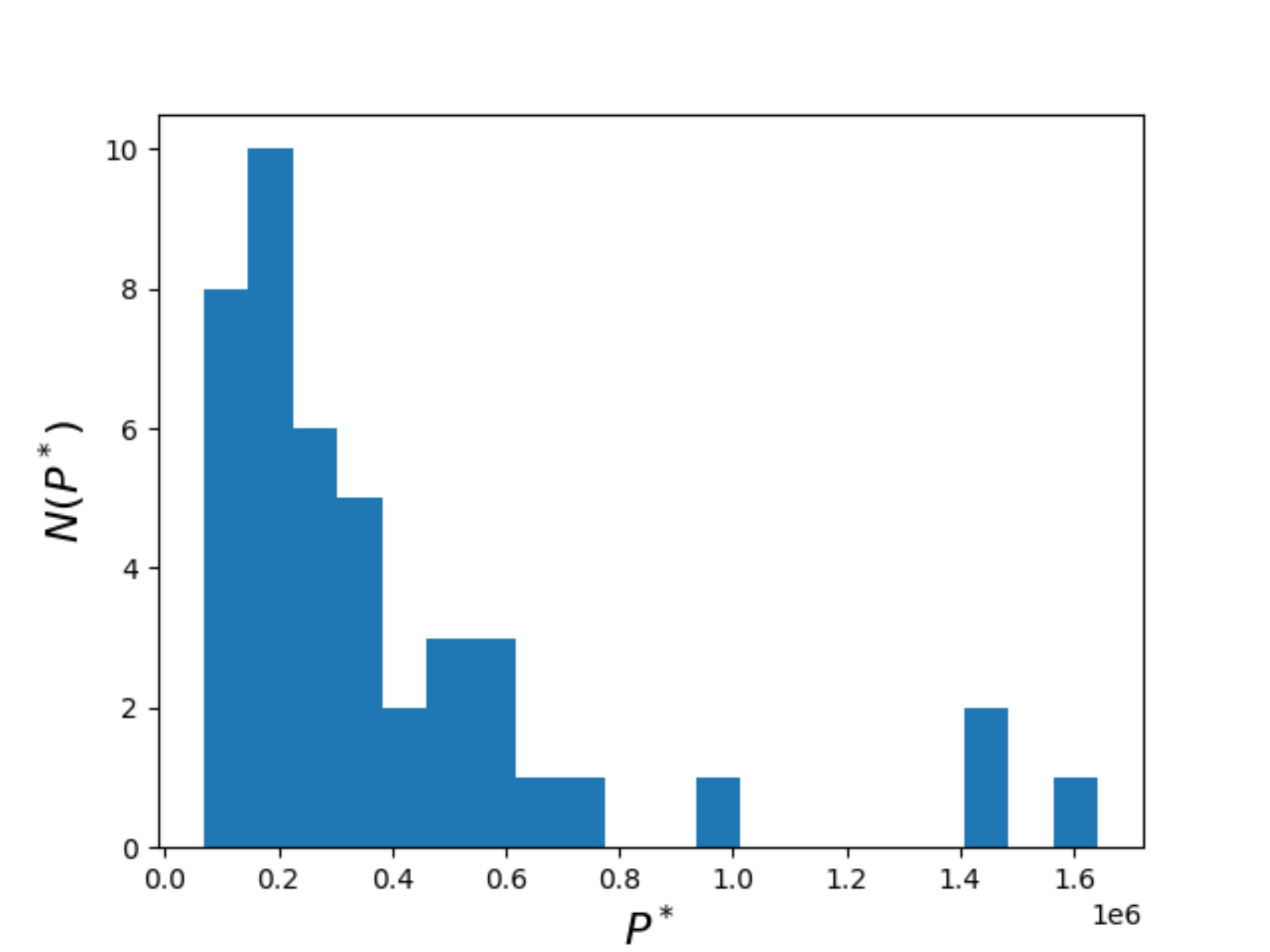} 
\includegraphics[scale=0.40]{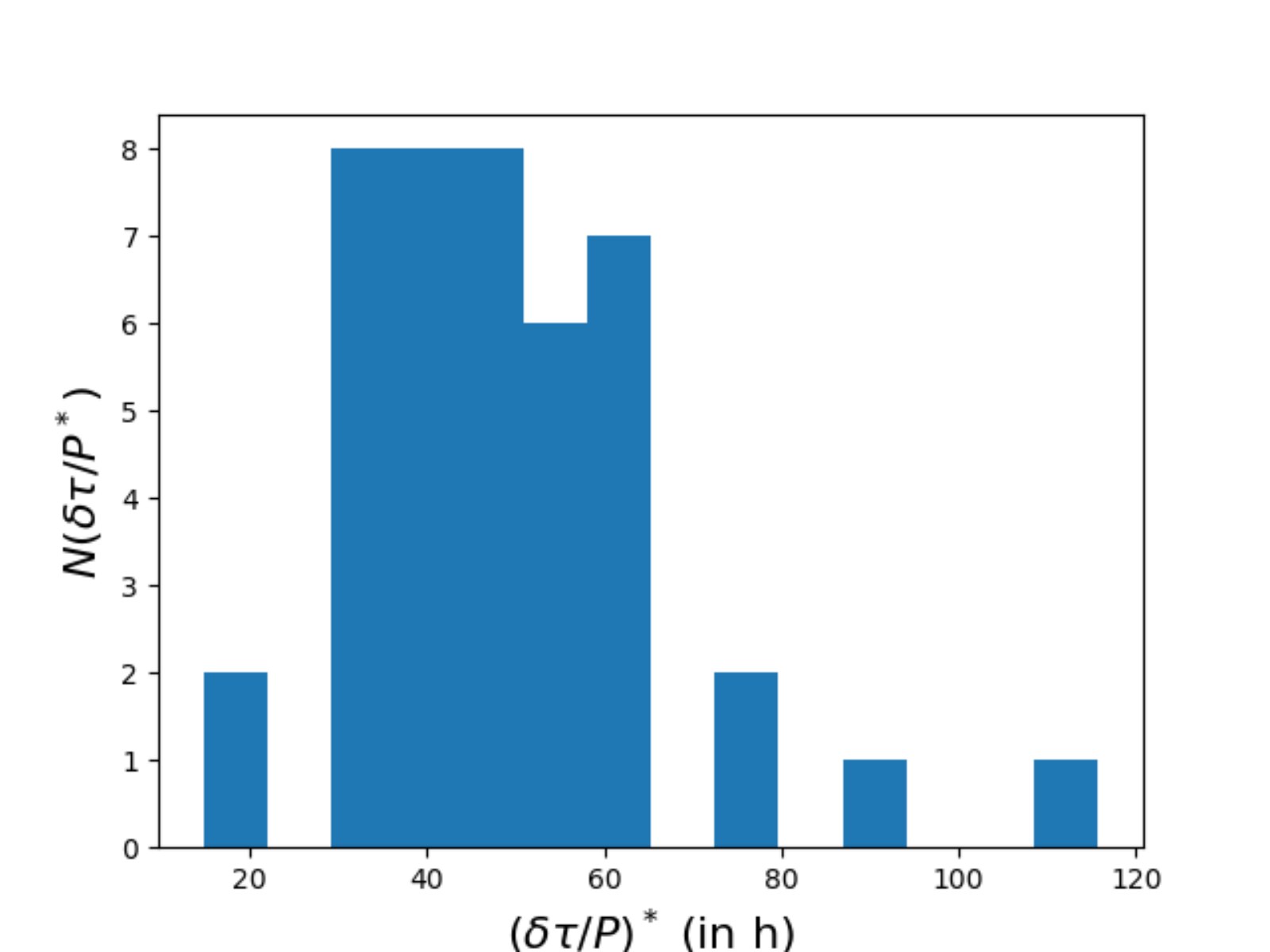} 
\caption{Empirical histograms for $T^*$, $P^*$ (in unit of million inhabitants) 
and $(\delta \tau/P)^*$ (in unit of hours). In particular the histogram for $(\delta \tau/P)^*$ shows that the changes of slope in type-2 cities appears approximately at the same value of about $40$ hours per year and per capita of congestion induced delay.}
\label{fig:rupture}
\end{figure}

\clearpage

\subsection{}
\centering{\bf Tables}
\begin{table}[htb!]
\centering
\begin{tabular}{|c|c|}
\hline
Cities & $\beta$ \\
\hline
Bakersfield CA & $3.1962881588682959$ \\
Baltimore MD&2.1270202500885378\\
Beaumont TX&4.3121047081974959\\
Brownsville TX&3.0747686515178829\\
Buffalo NY&4.2755491010100437\\
Corpus Christi TX&2.2675957429714364\\
Denver-Aurora CO&1.9671893783052652\\
Fresno CA&1.7450280827103009\\
Hartford CT&4.2648453780225291\\
Laredo TX&2.7791991465460999\\
Los Angeles-Long Beach-Anaheim CA&1.6285956803032693\\
Madison WI&2.0708101241687573\\
Miami FL&1.7546710159878378\\
New York-Newark NY-NJ-CT&2.4298069942492848\\
Phoenix-Mesa AZ&1.1968694416385248\\
Poughkeepsie-Newburgh NY-NJ&2.2768614183888238\\
Riverside-San Bernardino CA&2.2120810624454421\\
Rochester NY&2.7673045447707381\\
Sacramento CA&1.8017150060867237\\
Salt Lake City-West Valley City UT&2.9767035187758353\\
San Diego CA&2.0458776650851993\\
San Francisco-Oakland CA&1.6629530945958924\\
San Juan PR&3.1783030977078384\\
Sarasota-Bradenton FL&1.4488835644752018\\
Seattle WA&1.6121623183491192\\
Springfield MA-CT&2.6800144800260295\\
Stockton CA&2.1574496507830698\\
Tampa-St. Petersburg FL&1.6447950508957183\\
Toledo OH-MI&3.3688674059068782\\
Tulsa OK&2.7543394208073106\\
Virginia Beach VA&2.6761368087062629\\
\hline
\end{tabular}
\caption{Table for the exponent $\beta$ for type-1 cities.}
\label{tab:tab1}
\end{table}

\begin{table}[!h]
\centering
\begin{tabular}{|c|c|c|}
\hline
Cities & $\beta_1$ & $\beta_2$ \\
\hline
Akron OH&13.326474795174038&0.87036942722175636\\
Albuquerque NM&2.5254001684820788&0.43854750637699613\\
Allentown PA-NJ&3.5421104701666195&-0.067986378704719463\\
Anchorage AK&4.4297963653591861&1.8793584427200676\\
Baton Rouge LA&11.668425227500549&1.8982765309794534\\
Birmingham AL&5.7127486536782648&1.0950883331666841\\
Boston MA-NH-RI&3.7914578273352344&0.77783357746260684\\
Cape Coral FL&2.0170904969164649&-0.026908122725563643\\
Charleston-North Charleston SC&2.4180178167624629&0.73588836271136682\\
Cincinnati OH-KY-IN&13.391227137876555&1.4320691394968488\\
Colorado Springs CO&3.5302691471058441&0.90526757676454483\\
Dayton OH&10.888618834160816&1.5156113313272581\\
El Paso TX-NM&2.9723440807376567&0.38670363662700824\\
Eugene OR&9.6712863372110895&1.6336801378157653\\
Grand Rapids MI&4.9081888328711596&1.5063662641986844\\
Greensboro NC&4.2842149874033737&1.6655920377612741\\
Honolulu HI&2.6651812393270475&0.77842406987398327\\
Indianapolis IN&2.6903465538846123&1.090315534230037\\
Jackson MS&1.9548784600955464&4.7321108315657057\\
Kansas City MO-KS&7.5792226148887991&1.5218080339581797\\
Knoxville TN&5.3636780273848892&0.93514977131710486\\
Lancaster-Palmdale CA&1.2247810270154447&2.9739214893281631\\
Little Rock AR&4.7652094890503749&1.4888932143083471\\
Louisville-Jefferson County KY-IN&3.5668794174079124&0.9138483476564403\\
McAllen TX&2.4226867170036792&0.97162477887304632\\
Memphis TN-MS-AR&5.4918835943332702&1.8984236927813019\\
Milwaukee WI&7.421602499615239&1.4476849327745018\\
New Haven CT&6.6974370590880659&1.7283783197405369\\
Oklahoma City OK&3.5828355912773997&0.96908077750423693\\
Omaha NE-IA&5.2900061281968238&1.8104815755057855\\
Oxnard CA&4.2415757905874498&0.76412979056781882\\
Pensacola FL-AL&3.4069992036124965&1.3307657898250249\\
Philadelphia PA-NJ-DE-MD&8.8036825555975149&1.864839051262341\\
Pittsburgh PA&12.933233804630934&1.6230810430641514\\
Providence RI-MA&6.0658243195175512&1.8788870804188385\\
Raleigh NC&3.7254322679051999&1.5426528575241691\\
Salem OR&10.698703161001736&1.6924113979752105\\
San Antonio TX&2.1755450079365595&0.93642743110819815\\
San Jose CA&6.1181308990910814&2.0497802182301532\\
St. Louis MO-IL&4.0680015596238288&1.3568682133833097\\
Washington DC-VA-MD&2.4099008223841625&0.57016625283470113\\
Wichita KS&3.713495596552173&1.4821634370695835\\
Winston-Salem NC&2.003627578146336&-0.11462030190236838\\
\hline
\end{tabular}
\caption{Table for the exponents $\beta_1$ and $\beta_2$ for type-2 cities.}
\label{tab:tab2}
\end{table}

\end{document}